# NaCl crystal from salt solution with far below saturated concentration under ambient condition

Guosheng Shi[1†], Liang Chen[1,2,3†], Yizhou Yang[1†], Deyuan Li[1], Zhe qian[1,2], Shanshan Liang[1], Long Yan[1], LuHua Li[4], Minghong Wu[2*], and Haiping Fang[1*]

[1]Division of Interfacial Water and Key Laboratory of Interfacial Physics and Technology , Shanghai Institute of Applied Physics, Chinese Academy of Sciences, Shanghai 201800, China.
[2]Shanghai Applied Radiation Institute, Shanghai University, Shanghai 200444, China.
[3]Zhejiang Provincial Key Laboratory of Chemical Utilization of Forestry Biomass, Zhejiang A&F University, Lin'an, Zhejiang 311300, China
[4]Institute for Frontier Materials (IFM), Deakin University, Waurn Ponds, Victoria 3216, Australia
[†] These authors contributed equally to this work
*Corresponding author. E-mail: fanghaiping@sinap.ac.cn (H.-P. F.); mhwu@mail.shu.edu.cn (M.-H. W.)

**Abstract**

**Under ambient conditions, we directly observed NaCl crystals experimentally in the rGO membranes soaked in the salt solution with concentration below and far below the saturated concentration. Moreover, in most probability, the NaCl crystals show stoichiometries behavior. We attribute this unexpected crystallization to the cation-π interactions between the ions and the aromatic rings of the rGO.**

Under ambient conditions, NaCl crystal can be form in the solution with saturated concentration on a substrate without matching its lattice. On the unsaturated solution crystallizations has been reported by using the laser-induced method for Phenylalanine[1], glycine[2] and $NaClO_3$[3], as well as on a similar substrate with matching lattice for $PbSO_4$ and $SrSO_4$[4]. Here, under ambient conditions, we show a direct experimental observation of NaCl crystals in the rGO membranes soaked in the salt solution with concentration below and far below the saturated concentration. Moreover, in most probability, the NaCl crystals show abnormal stoichiometries behavior. We note that sodium chlorides with different stoichiometris have been reported to be thermodynamic stable only at extreme conditions, such as high pressure[5].

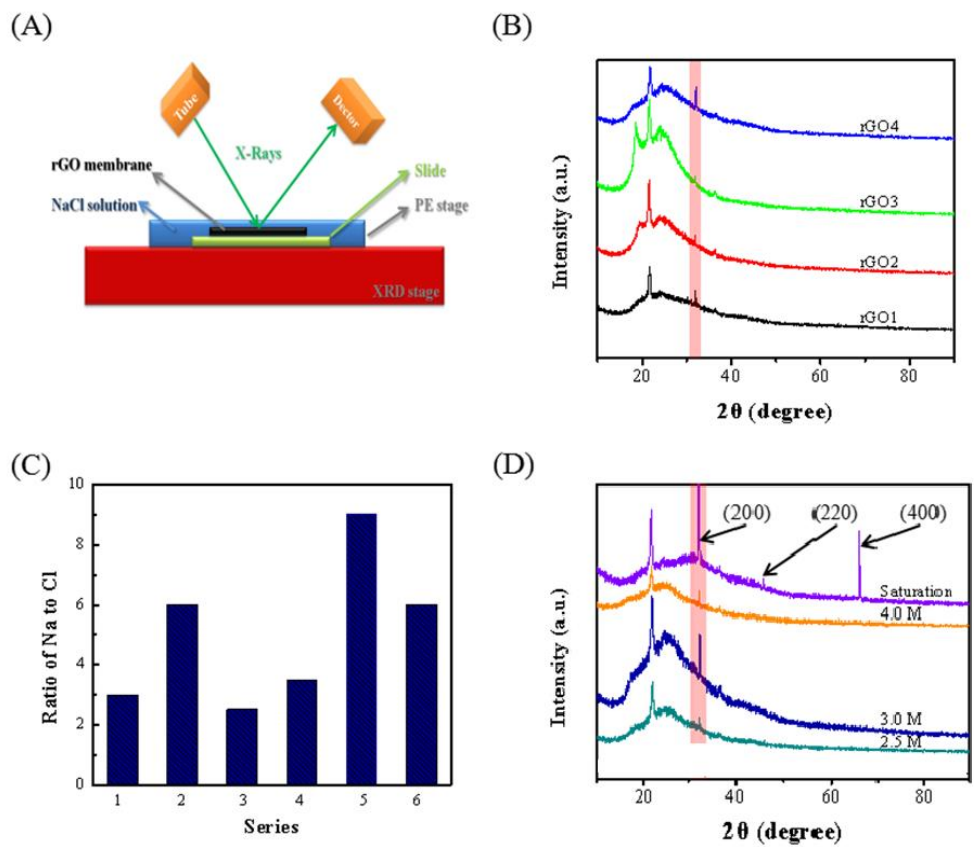

**Fig. 1. Two-dimensional NaCl crystals of stoichiometries from salt solution with far below saturated concentration.** (A) Schematic picture of experimental set-up. (B) XRD patterns of GO membranes in the solution (3.0M). (C) Ratios of Na and Cl of the crystals. (D) XRD patterns of NaCl crystals obtained from the rGO membranes immersed in different solution of 2.5 M, 3.0 M, 4.0 M and saturated concentration of 5.1 M.

Freestanding rGO membranes were prepared from an rGO suspension via the drop-casting method (details in Supplementary Information PS1). These membranes were then immersed for half hour in 3.0 mol/L (M) NaCl solutions in a PE bag (Fig. 1A). Next, the membranes with salt solution in the PE bag were analyzed by X-ray diffraction (XRD) (Fig. S1). There are many Bragg peaks at diffraction angles (2θ)

(Fig. 1B). The peaks of ~ 19, 22 and 25 degrees come from the rGO membranes, PE bags as well as slides, and the peaks of ~ 37 degree come from the PE bags and slides (Fig. S1B-E in SI). Only the peak of ~ 32 degree is a new Bragg peak, which is close to the value of 2θ for the {200} surface of NaCl crystal. We note that a single XRD peak of the surface is the typical characteristic of two-dimensional materials[6]. Considering that the saturated concentration of NaCl in aqueous solution is ~ 5.1 M, these results present that the existence of two-dimensional crystal in the pure aqueous NaCl solution with concentration far below the saturated concentration at ambient condition.

The NaCl crystal behaviors in rGO membranes can be observed for the outside solution with a wide range of concentrations over 2.5 M. The XRD results show that there are still clear new Bragg peaks at diffraction angle (2θ) ~ 32 degrees (Fig. 1D) at NaCl concentrations of 2.5 and 4.0 M. In contrast, the Bragg peaks of the general NaCl crystal can all be clearly seen when the solution has a supersaturated concentration, indicating that a general NaCl crystal nucleation occurs at this concentration. We note that the new peaks in the systems with concentration in the range of 2.5M to 4.0 M have clearly shifts to the (right) peaks of {200} surfaces of the general NaCl crystal (Fig. S1F in SI), indicating that the two-dimensional NaCl crystal has the {200} surface slightly different from the {200} surface of the general NaCl crystal.

We used energy dispersive X-ray spectra (EDS) to measure the element contents in the rGO sheets (see details in SI). The membranes employed in EDS experiments were obtained by first soaking with salt solution, then removal of the free solution by centrifugation and drying at 70 °C for 2 hours. We found that the ratios of the Na and Cl elements ranged from 2.5 to 9 (Fig. 1C). This is unexpected again since the ratios of the Na and Cl elements in the solution is 1:1 and at ambient conditions NaCl with the ratios of the Na and Cl elements of 1:1 is the only known stable compound in the Na-Cl system.

Why there is two-diemsional crystallization in the NaCl solution inside the rGO membrane soaked in the solution with far below saturated concentration. We attribute it to the ion-π interactions between the Na$^+$/Cl$^-$ and the aromatic rings of the rGO. Our previous MD simulations, incorporated with a force field of hydrated ion-π interaction, have shown the enrichment of Na$^+$/Cl$^-$ on the graphitic surface solvated in the aqueous NaCl solutions[7]. On the graphitic surface, the hydrated water on the ions may be separated due to the strong ion-π interactions, and the accumulated ions form crystals (see details in SI). Fig. 2 shows the stable two-dimensional crystal structures on the graphitic surfaces based on ab initio MD simulations. However, different from the general NaCl structures where the ratio of the Na and Cl elements is always 1:1, the ratios of the Na and Cl elements for these two-dimensional crystal structures can be 1:1 and 2:1. This is consistent with the experimental observation on the element contents in the rGO sheets.

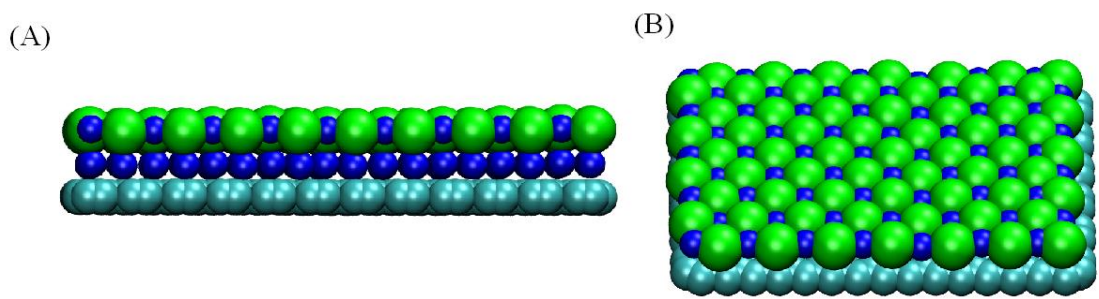

**Fig. 2 Density functional theory computations of the Na-Cl crystal structures on the graphitic surfaces.** The stable optimized geometries with the ratios of 2:1 (A) and 1:1 (B) of the Na and Cl elements on the graphitic surfaces. Spheres in cyan, blue and yellow represent C atoms, Na$^+$ and Cl$^-$.

In summary, we observed NaCl crystals experimentally in the rGO membranes soaked in the salt solution with concentration below and far below the saturated concentration under ambient conditions. Moreover, in most probability, the NaCl crystals show multi-stoichiometries behavior. We attribute this unexpected crystallization to that the cation-π interactions between the ions and the aromatic rings of the rGO. The finding enriches our understanding of the crystallization as well as the structures of the crystals.

**References and Notes:**

1  Yuyama, K.-i. et al. Two-Dimensional Growth Rate Control of l-Phenylalanine Crystal by Laser Trapping in Unsaturated Aqueous Solution. *Cry. Growth Des.* **16**, 953-960 (2016).

2  Yuyama, K.-i. et al. Selective Fabrication of α- and γ-Polymorphs of Glycine by Intense Polarized Continuous Wave Laser Beams. *Cry. Growth Des* **12,** 2427 (2012).

3  Niinomi, H. *et al.* Plasmonic Heating-Assisted Laser-Induced Crystallization from a NaClO3 Unsaturated Mother Solution. *Cry. Growth Des* **17**, 809-818 (2017).

4  Murdaugh, A. E. *et al.* Two-Dimensional Crystal Growth from Undersaturated Solutions. *Langmuir* **23**, 5852-5856 (2007).

5  Zhang, W. *et al.* Unexpected Stable Stoichiometries of Sodium Chlorides. *Science* **342,** 1502-1505 (2013).

6  Lin, Z. *et al.* Scalable solution-phase epitaxial growth of symmetry-mismatched heterostructures on two-dimensional crystal soft template. *Sci. Adv.* **2**, e1600993 (2016).

7  Shi, G. *et al.* Ion enrichment on the hydrophobic carbon-based surface in aqueous salt solutions due to cation-π interactions. *Sci. Rep.*, **3**, 3436 (2013).

**Acknowledgments:** The supports from NSFC (Nos. 11574339 and 11404361) and Supercomputer Center of CAS, BL16B1 and BL14W1 beamline at SSRF are acknowledged.

**Author contributions:** H.-P. F. had the idea to observe NaCl crystals in the rGO membranes in the salt solution with concentration below the saturated concentration based on the ion-π interactions. H.-P. F., M.-H. W., G.-S. S. and L. Y. conceived and designed the experiments and simulations. L. C., D.-Y. L. and Z. Q. performed the experiments. G.-S. S. and Y.-Z. Y. performed the simulations. G.-S. S., L. C., H.-P. F., S.-S. L., L. Y. and L.-H. L. analysed the data, G.-S. S., H.-P. F., S.-S. L. and M.-H. W. co-wrote the paper. All authors discussed the results and commented on the manuscript.